\documentclass[preprint2]{aastex6}

\usepackage{graphicx,epsfig,fancyhdr,psfig,rotating,amsmath,epsf,txfonts,natbib,epstopdf,multirow}
\def \kms {{\rm km\;s$^{-1}$}}

\def \arcsec {$^{''}$}

\graphicspath{{./figs/}} 
\def \halpha {{H$\alpha$}}

\begin{document}
\title{A magnetic reconnection event in the solar atmosphere driven by relaxation of a twisted arch filament system}
\author{
\sc{Zhenghua Huang\altaffilmark{1}, Chaozhou Mou\altaffilmark{1}, Hui Fu\altaffilmark{1}, Linhua Deng\altaffilmark{2}, Bo Li\altaffilmark{1}, Lidong Xia\altaffilmark{1}}
}
\altaffiltext{1}{Shandong Provincial Key Laboratory of Optical Astronomy and Solar-Terrestrial Environment, Institute of Space Sciences, Shandong University, Weihai, 264209 Shandong, China; {\it z.huang@sdu.edu.cn}}
\altaffiltext{2}{Yunnan Observatories, Chinese Academy of Sciences, Kunming, 650216, China}

\begin{abstract}
We present high-resolution observations of a magnetic reconnection event in the solar atmosphere taken with the New Vacuum Solar Telescope, AIA and HMI.
The reconnection event occurred between the threads of a twisted arch filament system (AFS) and coronal loops.
Our observations reveal that the relaxation of the twisted AFS drives some of its threads to encounter the coronal loops, providing inflows of the reconnection.
The reconnection is evidenced by flared X-shape features in the AIA images, a current-sheet-like feature  apparently connecting post-reconnection loops in the \halpha$+$1\,\AA\ images, small-scale magnetic cancellation in the HMI magnetograms and 
flows with speeds of 40--80\,\kms\ along the coronal loops.
The post-reconnection coronal loops seen in AIA\,94\,\AA\ passband appear to remain bright for a relatively long time, 
suggesting that they have been heated and/or filled up by dense plasmas previously stored in the AFS threads.
Our observations suggest that the twisted magnetic system could release its free magnetic energy into the upper solar atmosphere through reconnection processes.
While the plasma pressure in the reconnecting flux tubes are significantly different, the reconfiguration of field lines could result in transferring of mass among them and induce heating therein.
\end{abstract}

\maketitle

\section{Introduction}
\label{sect_intro}
Magnetic reconnection, which can convert magnetic energy to heat and kinetic energy, is a fundamental process in the conductive plasma.
It has been thought to be the major cause of a various of active phenomena on the sun, from small-scale explosive events\,\citep[e.g.][]{1991JGR....96.9399D,Innes1997Nature,Huang2014ApJ,2015ApJ...810...46H,Huang2017mnras}, Ellerman and IRIS bombs\,\citep[e.g.][]{2013ApJ...779..125N,Peter2014Sci,2016ApJ...824...96T,2017ApJ...836...52Z}, bright points\,\citep[e.g.][]{1994ApJ...427..459P,2012ApJ...746...19Z,2012A&A...548A..62H,2016ApJ...818....9M} and jets\,\citep[e.g.][]{1999ApJ...510..485A,Tian2014Sci,Sterling2015Nature,Tian2017ApJ}, to large scale flares and CMEs\,\citep[e.g.][]{2000JGR...105.2375L,2016ApJ...820L..37C,2016ApJ...823..136Z,2016ApJ...832..195N}.
In the solar observations,
many evidences of magnetic reconnection, including plasma inflows/outflows, current sheet structures, shocks and plasma ejections and heatings, have been frequently reported.
However,
it is impossible to observe the full processes of magnetic reconnection on the sun directly because of the limitations in the remote-sensing data.
In the upper solar atmosphere, where the plasma beta is low, 
the geometry of magnetic field might be traced via the evolution of plasma structures,
and thus allow one to infer the processes of magnetic reconnection. More details of magnetic reconnection and its applications can be found in the textbooks written by \citet{priest2000book} and \citet{2014masu.book.....P}.

\par
Thanks to the high temporal and spatial resolution data obtained in the past decade,
it has reported on a few sets of observations that directly verify the pictures of magnetic reconnection in the solar atmosphere.
Using AIA and RHESSI data, \citet{su2013} present the processes of magnetic reconnection in the corona occurred during a modest flare. In their case, the magnetic reconnection occurred 
when a set of coronal loops were moving toward another set of loops and then creating an X-shape geometry of coronal structures. 
Newly formed coronal loops and sources of hard X-ray emissions were observed after the disconnection of the X-shape structure, which demonstrated the occurrence of magnetic reconnection.
Using \halpha\ images taken by the New Vacuum Solar Telescope (NVST),
\citet{2015ApJ...798L..11Y} reported observations of magnetic reconnection in the chromosphere between two sets of anti-parallel filament threads, which also show clear responses in AIA EUV images.
They observed that approaching of anti-parallel loops to form an X-shape geometry is the overture of the magnetic reconnection.
The plasma inflows/outflows of the reconnection processes were also clearly present in their data.
With AIA and the twin spacecrafts of STEREO, \citet{Sun2015NatCo} imaged an eruptive event in the solar atmosphere at three different viewing angles.
They observed that the reconnection processes include gradually approaching of two groups of opposite directed loops, generating of a quasi-separator and forming of new loops.
Their observations are thought to be consistent with the theoretical three-dimensional configuration of magnetic reconnection.
In another case, \citet{Li2016NatPh} reported AIA observations of a magnetic reconnection event between an eruptive solar filament and nearby coronal loops.
In their case, an X-shape structure was observed while the eruptive filaments ran into the coronal loops, and then bright current-sheet-like features were formed and bi-directional plasma ejections were produced.
While analysing a solar filament eruption observed by NVST and SDO, \citet{Xue2016NatCo} observed the processes of magnetic reconnections between the filament and chromospheric fibrils.
In their observations, the filaments and fibrils are approaching (inflows), and then cusp structures and elongated bright structure (current sheet) are forming.
The disconnection of the cusp structure forms new loops that are moving bi-directionally (outflows).

\par
Magnetic reconnection in the solar atmosphere has been confirmed by the previous observations,
but the causes of the inflows that directly drive magnetic reconnection are rarely observed.
In the present work, we present observations of a small-scale active phenomena in the solar atmosphere that caused by magnetic reconnection between threads of an arch filament system (AFS) and coronal loops.
In this particular case, we will see that the inflows of the magnetic reconnection is driven by relaxation of the twisted AFS, and the reconnecting system has been significantly flared up.

\begin{figure}[!ht]
\centering
\includegraphics[trim=1.2cm 1cm 0.5cm -0.5cm,width=\linewidth]{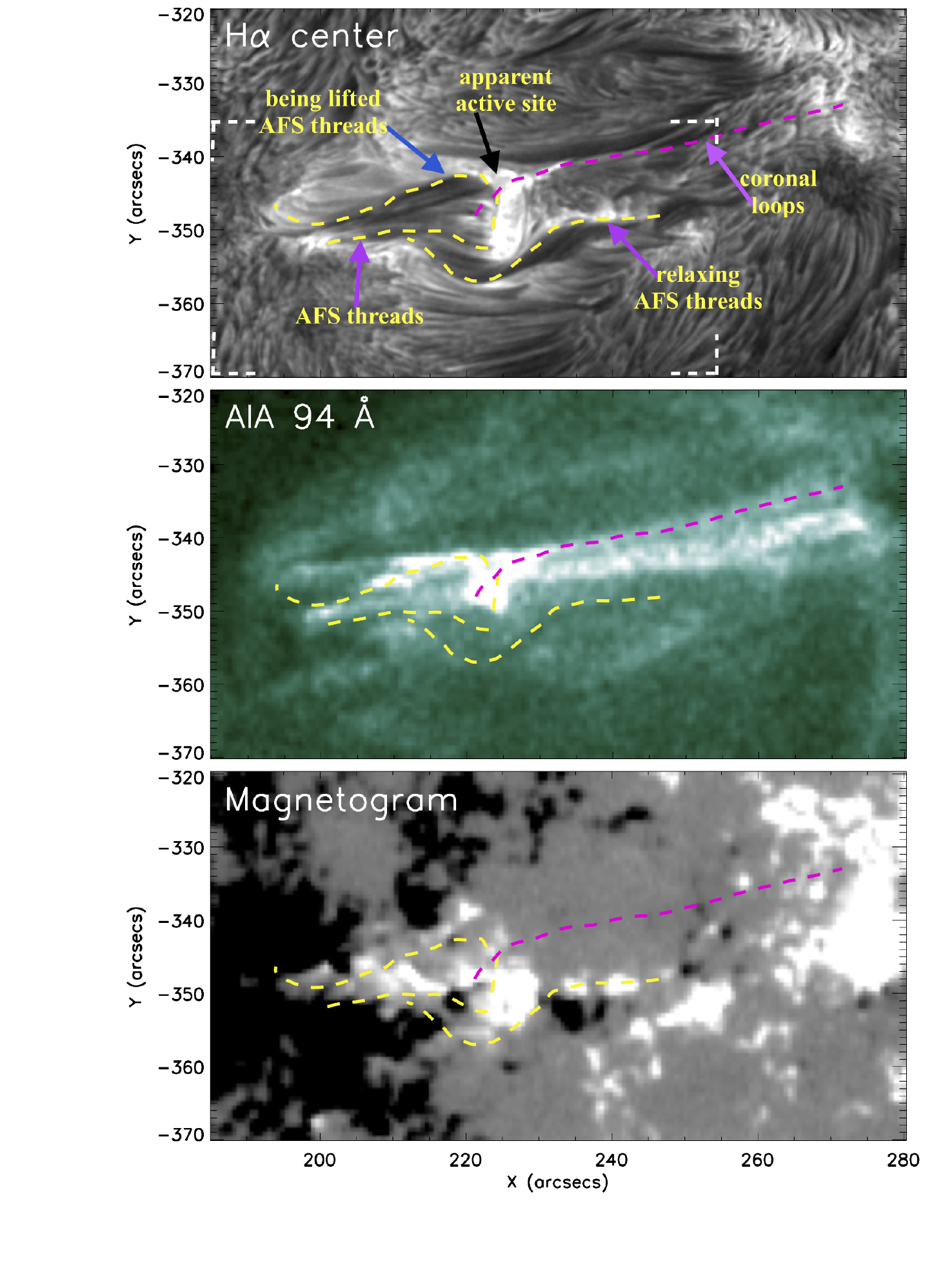}
\caption{The region seen with NVST \halpha\ center (top), AIA 94\,\AA\ (middle) and HMI line-of-sight magnetogram (bottom) at 07:12:28\,UT, 07:12:24\,UT and 07:12:33\,UT, respectively.
The yellow dashed lines trace a few threads of the AFS seen with NVST \halpha\ center.
The purple dashed lines denote the coronal loops seen with AIA 94\,\AA.
The HMI magnetogram is artificially saturated at $-$200\,G (black) and 200\,G (white).
The evolution of the region (X:185\arcsec$\sim$255\arcsec, Y:$-$370\arcsec$\sim$$-$335\arcsec) with corners marked by white dashed lines in the \halpha\ image is shown in Figure\,\ref{fig:flactivity}.
(An animation of the region seen in multiple passbands is given online.)
}
\label{fig:overview}
\end{figure}

\section{Observations}
Our observing campaign was carried out on September 25 2016 from 06:32\,UT to 07:52\,UT with a target of AR12597.
The data were obtained by the ground-based NVST\,\citep{2014RAA....14..705L} that is one of the main observing facilities in the Fuxian Lake Solar Observing stations operated by the Yunnan Astronomical Observatories in China, 
and the space-born Atmospheric Imaging Assembly\,\citep[AIA,][]{Lemen2012aia} and the Helioseismic and Magnetic Imager\,\citep[HMI,][]{Schou2012hmi} onboard the Solar Dynamics Observatory\,\citep[SDO,][]{Pesnell2012sdo}.
The NVST data include series of \halpha\ images taken at line centre and offband at $+$1.0\,\AA\ and $-$1.0\,\AA\ with a bandpass of 0.25\,\AA.
The cadence of each series is about 50\,seconds after being reconstructed with Speckle technics\,\citep[][and references therein]{2016NewA...49....8X}.
The stabilization of image series is carried out by a fast sub-pixel image registration algorithm provided by the instrument team\,\citep{Feng2012IEEE,2015RAA....15..569Y}.
The spatial sampling of the \halpha\ images is 0.17\arcsec per pixel.
The AIA data primarily presented here include series UV/EUV images taken with the passbands of 1600\,\AA, 304\,\AA, 171\,\AA, 193\,\AA, 131\,\AA\ and 94\,\AA.
The spatial resolution of the AIA data is 1.2\arcsec,
and the cadences are 24\,s for 1600\,\AA\ passband and 12\,s for the rest. 
To study the evolution of the magnetic features associated with the active event, the HMI line-of-sight magnetograms with a cadence of 45\,s and a spatial resolution of 1\arcsec are used.

\par
The images from different instruments and passbands have been aligned with a principle of best cross-correlation between images of two passbands with the closest representative temperatures.
The \halpha\ images have also been rotated to fit into the Helioprojective-Cartesian coordinate system as used in the AIA and HMI data.

\begin{figure*}[!ht]
\centering
\includegraphics[trim=0cm 0cm 0cm 0cm,width=\linewidth]{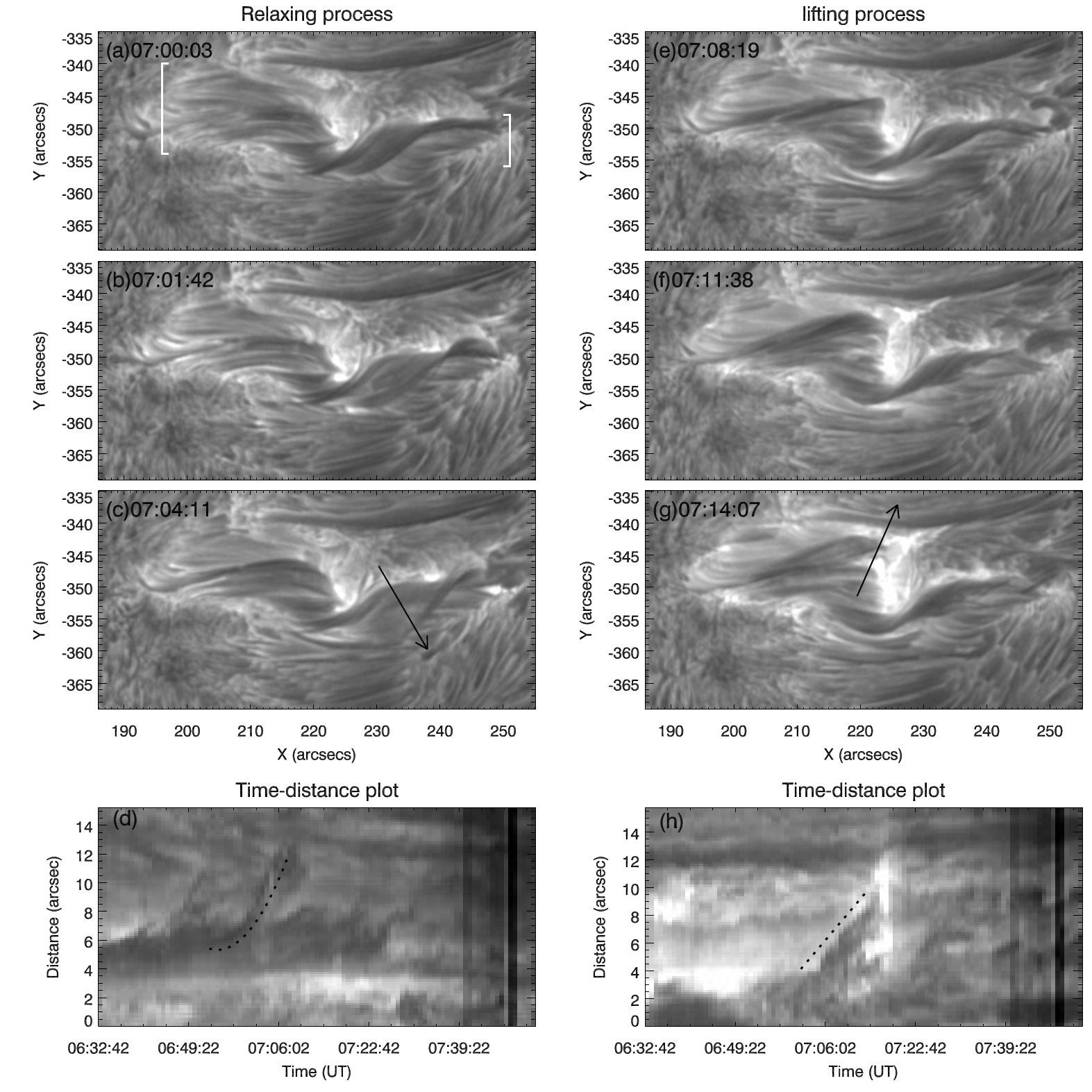}
\caption{Activities in the twisted AFS, whose east and west ends are denoted by white brackets in panel (a).
Panels (a--c): Transformation of the geometry of the AFS in the stage while its west portion was being untwined and moving southward (marked by the arrow in panel c).
Panel (d): The time-distance plot along the cut (marked by the arrow in panel c), recording the southward movement of the west AFS threads. The dark track indicated by dotted line is the major one associated with the reconnection event.
Panels (e--g): Transformation of the geometry of the AFS in the stage while its east portion was being lifted and moving northward (toward the pre-existed coronal loops, marked by the arrow in panel g).
Panel (h): The time-distance plot along the cut (marked by the arrow in panel g), giving the northward movement of the east AFS threads. The dark track below the indicated dotted line is the major one providing inflows to the magnetic reconnection.
(An animation is given online.)
}
\label{fig:flactivity}
\end{figure*}

\begin{figure*}[!ht]
\centering
\includegraphics[trim=0.5cm 1.5cm 1cm 0cm,width=\linewidth]{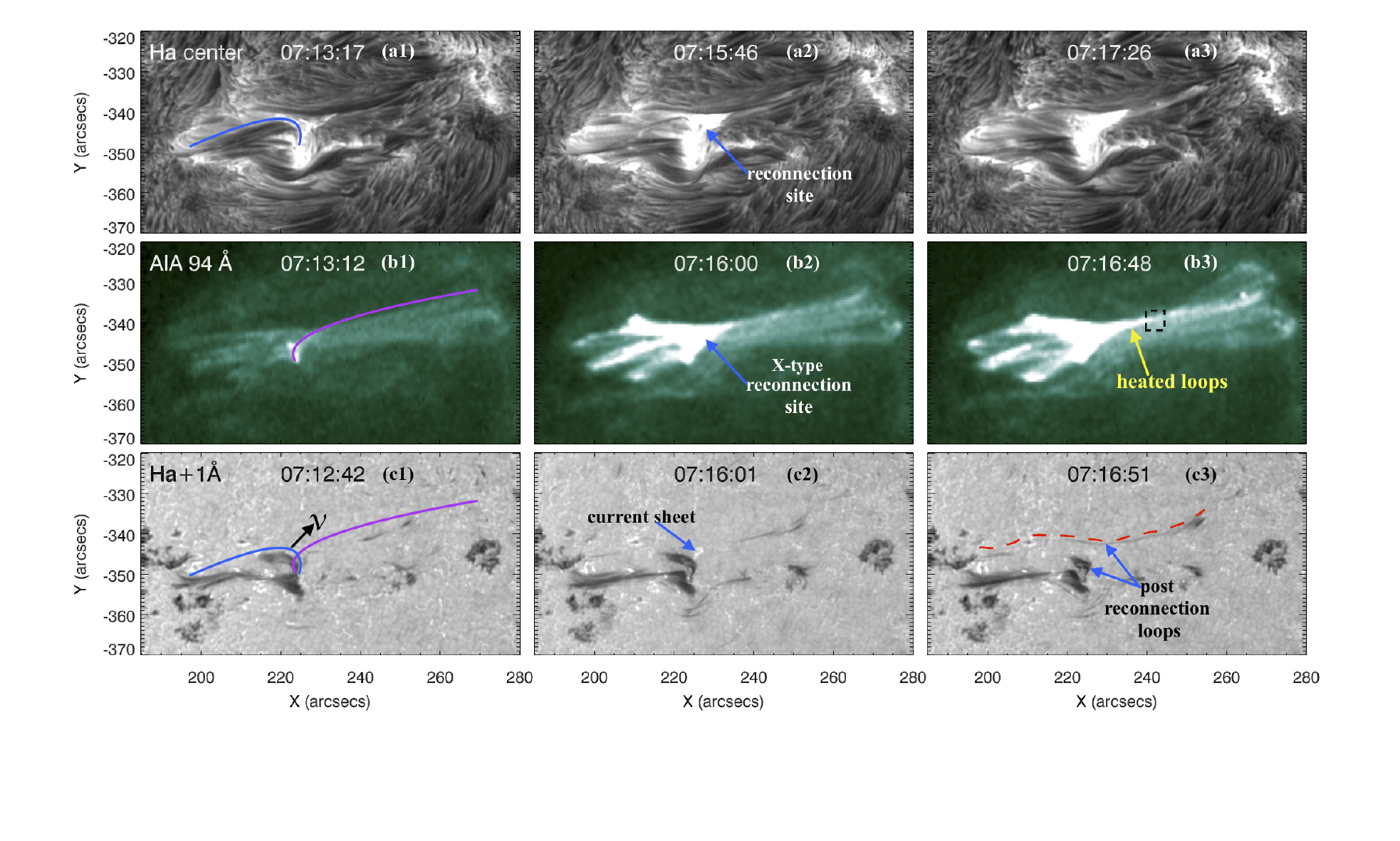}
\includegraphics[trim=0cm 0cm 0cm 0cm,width=\linewidth]{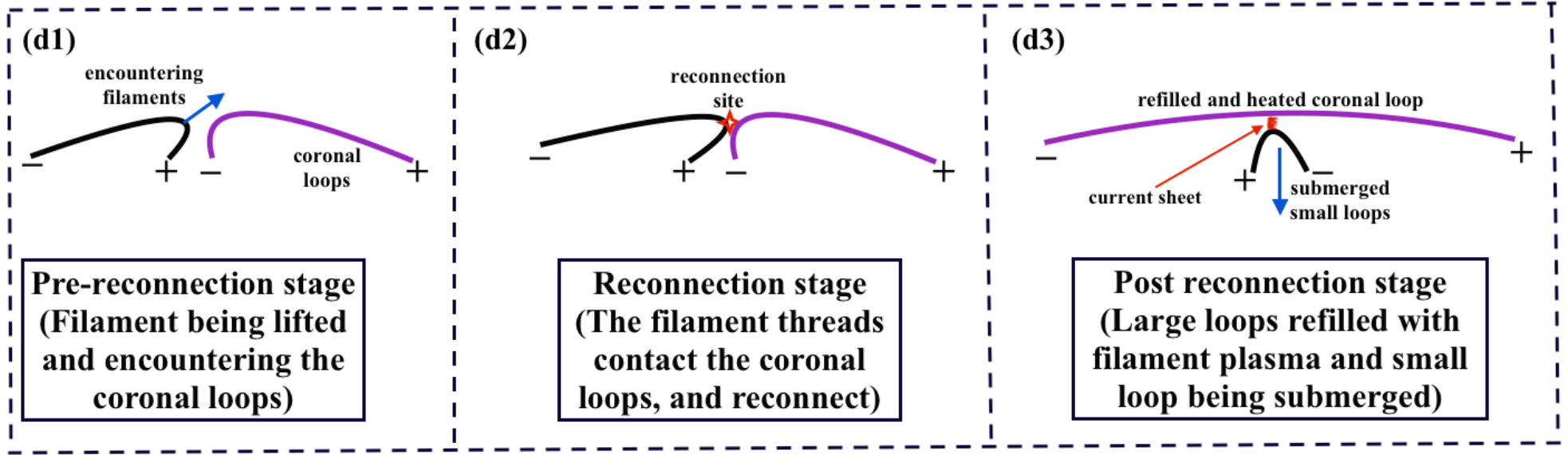}
\caption{ The transformations of the connectivities in the region of interest seen in NVST \halpha\ line centre (a1--a3), AIA 94\,\AA\ (b1--b3) and NVST \halpha$+$1\,\AA\ (c1--c3).
The images show the connectivities just before the reconnection (a1, b1, c1), during the reconnection (a2,b2,c2) and slightly after (a3, b3, c3).
In the left column, the AFS threads that are being lifted are denoted by the blue line in panel a1 (the lifting process is better seen in the online animation associated with Figure\,\ref{fig:overview}), the coronal loops are marked by the purple line in panel b1, and both are illustrated on panel c1 with the black arrow indicates the northward motion of the AFS threads.
The remarkable patterns observed during the reconnection are denoted in the images.
The region enclosed by the black dashed lines in the AIA 94\,\AA\ image at 07:16:48\,UT is used to produce light-curves in Figure\,\ref{fig:prflow}.
d1--d3: A cartoon showing the processes of the magnetic reconnection as observed.
}
\label{fig:recprocess}
\end{figure*}

\begin{figure}[!ht]
\centering
\includegraphics[trim=0cm 1cm 0cm 0cm,width=\linewidth]{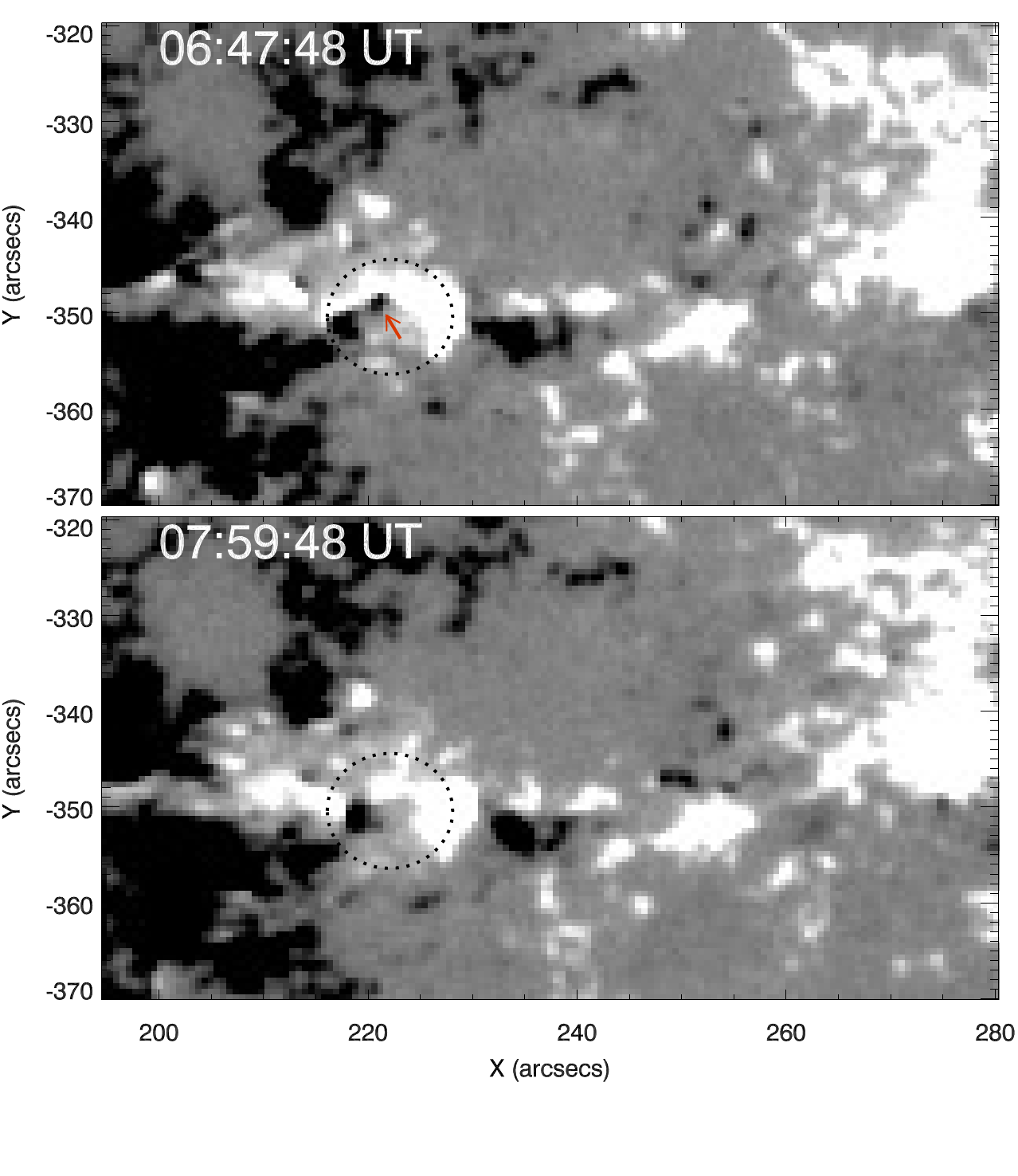}
\caption{The evolution of the HMI magnetograms in the inferred region of the magnetic reconnection, artificially saturated at $-$200\,G (black) and 200\,G (white). The place where the negative fluxes are clearly decreased is enclosed by the circular dotted lines. The arrow on the top panel denotes the negative feature that has been cancelled as seen in the bottom panel.
(An animation of the HMI magnetogram is given online.)
}
\label{fig:magevl}
\end{figure}

\begin{figure*}[!ht]
\centering
\includegraphics[trim=0cm 0.2cm 0cm 0cm,width=0.9\linewidth]{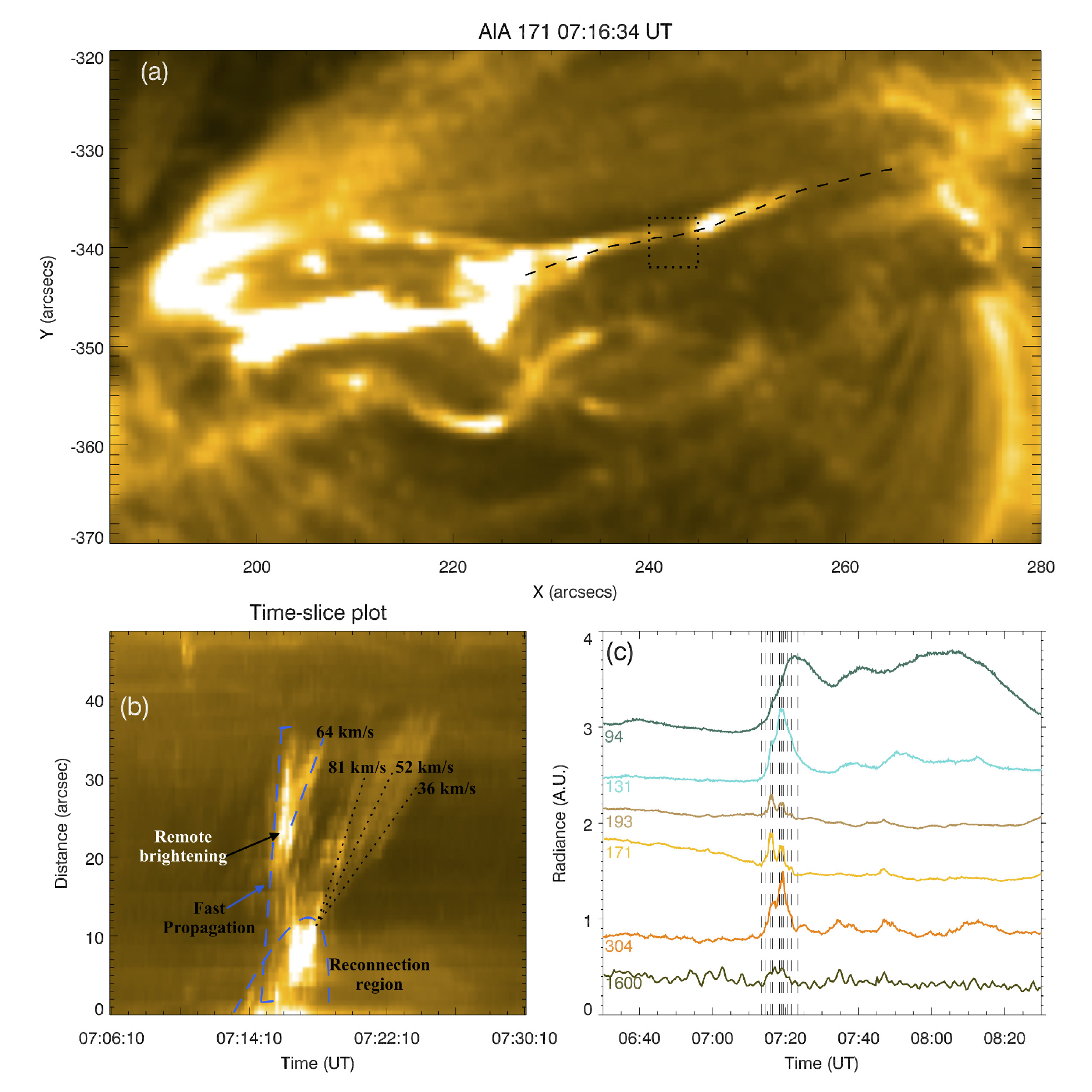}
\caption{(a): The studied region seen in AIA 171\,\AA.
The dashed line denotes the track of the flows along the coronal loops resulted from the reconnection event.
The dotted lines (square) enclose the region (as same as that in Figure.\,\ref{fig:recprocess}), from which the light-curves (panel c) were obtained.
(b): Time-distance plot obtained along the dashed line in panel (a),
recording the tracks of the apparent flows along the coronal loops.
Some representative features are denoted (see discussion in the text).
(c): The light-curves of the region enclosed by dotted line in panel (a) taken with different AIA channels.
The dashed lines are given for reference.
}
\label{fig:prflow}
\end{figure*}

\section{Results and discussions}
\label{sect_overview}
The general connectivities in the region is shown in Figure\,\ref{fig:overview}.
In \halpha\ images, we clearly observe a twisted arch filament system\,\citep[AFS,][]{1967SoPh....2..451B,1972SoPh...26..130F,2017SPD....4840604S}.
While following the long-term evolution of magnetic features in the region (see the online animation associated with Figure\,\ref{fig:magevl}), we found that this AFS is possibly formed by flux emergence with rotating motions in the footpoint polarities.
The AFS has a serpentine geometry, which is consistent with common emerging flux tubes\,\citep[see e.g.,][]{2004ApJ...614.1099P,2009ApJ...701.1911P,2014SSRv..186..227S}.
The east ends of the AFS threads appear rooting in a large patch of negative polarity,
while the west ends are positive polarities at various locations (see the bottom panel of Figure\,\ref{fig:overview}).
One set of the AFS threads (their west ends) is rooted in a patch of positive polarity at the location at about solar\_x=225\arcsec\ and solar\_y=-345\arcsec.
These AFS threads will be lifted by the untwined motion of the system (see Section\,\ref{sect_driver}).
This patch of positive polarity is also surrounding a small patch of negative polarity that is host of the east footpoints of coronal loops seen in the AIA\,94\,\AA\ passband.
The west ends of the coronal loops is situated in a large patch of positive polarity.
The magnetic reconnection event appear to occur in this region between the being lifted AFS threads and the coronal loops, and
the reconnection processes will be seen in the following descriptions.
Some threads of the coronal loops can be resolved earlier than 06:30\,UT and best visible in the AIA\,94\,\AA\ passband.
However, the threads involved in the active event (purple lines in Figure\,\ref{fig:overview}) appear to be obscured in the background emission at the early stage, and could be resolved in the AIA 94\,\AA\ images only after 07:12\,UT (see the online animation).

\par
The activities started at $\sim$07:00\,UT, while the west portion of the twisted AFS threads were apparently being untwined.
The untwisting motion progresses southward, 
and it results in northward motion in the east portion of AFS.
The northward motion forces the AFS threads moving toward the pre-existed coronal loops. 
The flaring activity is observed around 07:15\,UT when the AFS threads were encountering the coronal loops.
This indicates occurrences of magnetic reconnection between the AFS threads and the coronal loops.
The resulted flaring hot loops last for relatively-long period of time in the AIA 94\,\AA\ images.
The full evolution of the event can be viewed in the online animation attached to Figure\,\ref{fig:overview} and more details of the analyses are given in the following sections.

\subsection{Relaxation of the twisted AFS and inflows of the reconnection}
\label{sect_driver}
In this section, we focus on the untwined activities of the twisted AFS occurring at the early stage of the event.
Such untwined motions appear to drive the AFS threads moving toward the pre-existed coronal loops and thus provide the inflows of the magnetic reconnection.
The activities of the AFS are shown in Figure\,\ref{fig:flactivity} and the online animation.
The high-resolution \halpha\ images reveal AFS threads in very fine scale ($\sim$0.5\arcsec\ in width).
Well before the beginning of the AFS activity, we can see that the AFS threads are winding each other and forming a twisted AFS (Figure\,\ref{fig:flactivity}a).

\par
Starting from about 06:40\,UT, threads in the west portion of the AFS were intermittently untwined and moving southward (Figures\,\ref{fig:flactivity}a--c). 
The most clear untwined activities start from $\sim$06:54\,UT and can be clearly seen after 06:58\,UT.
From the time-distance image, we found that this rapid untwined motion has an apparent acceleration of $\sim$15\,m\,s$^{-2}$ along the cut (Figure\,\ref{fig:flactivity}d).
The acceleration could be results of the instability causing the untwining motion and/or the projection effects of the untwining circular motion.
The present observations, however, do not allow to distinguish the two processes.

\par
After $\sim$07:00\,UT, we observe that the threads in the east portion of the AFS appears to be lifted and starts to move northward (Figures\,\ref{fig:flactivity}e--g).
This northward motion appears to follow the untwined motion of the west portion of the AFS.
At $\sim$07:14\,UT, the site where these AFS threads contact the coronal loops (see Figure\,\ref{fig:overview}) start to flare up.
This suggests that the magnetic reconnection begins.
The speed of the northward lifted motion gives 5.5\,\kms\ measured from the time-distance image (Figure\,\ref{fig:flactivity}h).
This could be taken as the apparent velocity of the inflows of the magnetic reconnection.
The real velocity of the inflows should be larger if the projection effect is taken into account.

\subsection{Geometry of the magnetic reconnection}
\label{sect_geometry}
While following the evolution of the active event, we observed some clear patterns that are representative of the magnetic reconnection processes.
While the lifted AFS threads were clearly seen in \halpha\ line centre, this activity can also be observed in \halpha\  offbands (see the online animation associated with Figure\,\ref{fig:overview}).
The evolution of the event is shown in Figure\,\ref{fig:recprocess} and the animation 1.
In \halpha\ line center, beside the activity of the AFS, the flaring at the site where the AFS threads were encountering the coronal loops is also clearly observed (see Figure\,\ref{fig:recprocess} a1--a3).
This flaring can be seen in all AIA passbands (as the 94\,\AA\ images shown in Figure\,\ref{fig:recprocess} b1--b3 and a few others shown in the online animation).
The AIA images also reveal X-shape geometry of the event (Figures\,\ref{fig:overview}\&\ref{fig:recprocess}).
In the \halpha\ line center and all AIA passbands, we can also observe clear flows along the coronal loops, which will be investigated in next section.

\par
Viewed in \halpha$+$1\,\AA, dark features can be clearly seen in the footpoints of the threads (Figure\,\ref{fig:recprocess} c1--c3).
These features show motions toward the solar surface (see the online animation), 
which are possible the downflows in the AFS threads while they are being lifted.
While the reconnection initiated, the AFS threads remain dark in \halpha,
suggesting that the cool dense plasma in the AFS threads has not yet evacuated by the downflow.
While the inferred reconnection site was flaring ($\sim$07:16\,UT), we observed a bright elongated feature in \halpha$+$1\,\AA.
The size of the feature is about 3.5\arcsec in length and 0.5\arcsec\ in width.
At the two ends of the bright elongated feature
we observe a set of very large and a set of very small dark loop-like features (see Figure\,\ref{fig:recprocess} c3).
These loops are very likely the post reconnection loops, and thus the bright elongated feature could be representative of a current sheet forming in the reconnection.
The set of small loops seen in the \halpha$+$1\,\AA\ images/animations include threads with various sizes shrinking back toward the solar surface, which are downflows and/or submerging loops.

These observations could be simplified as a cartoon shown in Figure\,\ref{fig:recprocess} d1--d3.
The small-scale loops that are observed in the \halpha$+$1\,\AA\ images could be produced by the reconnection processes and eventually submerged.
The submergence of the small-scale loops could be observed as flux cancellations that is clearly seen in Figure\,\ref{fig:magevl}.
The larger-scale loops produced by the reconnection processes are hotter loops in the corona as viewed by AIA 94\,\AA\ channel.
In flux emerging region, large coronal loops could be formed through series of reconnection above bald-patches of the emerging serpentine flux tubes\,\citep[e.g.,][]{2004ApJ...601..530S,2004ApJ...614.1099P,2009ApJ...701.1911P,2017ApJ...836...52Z}.
While magnetic reconnection above bald-patches is assumed to be driven by line-tied motions,
the untwisting motion of the AFS observed here could be an alternative mechanism to produce larger coronal loops.
A question is whether the reconnecting AFS threads and coronal loops belong to the same emerging serpentine flux tubes.
The question cannot be fully answered with the present data, 
however, if they belong to the same emerging system, the coronal loops would require heating before the active event.

\par
Because the AFS threads could contain much denser plasma then the coronal loops, the pressure gradient in the newly formed large-scale loops could be important and drive a plasma flow therein.
The plasma density of the original coronal loops could increase according to this process and the newly-formed loops could be flaring up as observed in AIA 94\,\AA\ channel. 
The newly-formed loops could be then heated by the enthalpy carried in such a plasma flow.
Furthermore, the magnetized plasma will certainly flow along magnetic field lines\,\citep{He2017inprep}, while the component along the magnetic tension force defines the outflow of the magnetic reconnection. 
In this case, we clearly observe the plasma flow along the newly-formed loops and will investigate in next section.

\subsection{Flows and heating resulting from the reconnection}
The flows along the coronal loops can be seen in all AIA channels and \halpha\ line center (see the online animation).
The flows appear as multiple bright blobs running along the coronal loops.
The time-distance plot along the coronal loop taken with the AIA 171\,\AA\ channel (Figure\,\ref{fig:prflow}b) presents a few interesting patterns:
fast propagating ``flows'', slow flows with multiple speeds and remote brightening.
The fast propagating ``flows'' appear to originate in the reconnection region.
their apparent speeds exceed 1000\,\kms\ that cannot be accurately measured with the current data,
and it remains puzzling whether these are shock flows or sudden appearance of coronal loops that were previously obscured by overlying structures.
The apparent velocities of the slow flows measured from the three tracks of the multiple blobs give 36\kms, 52\kms\ and 81\,\kms.
This speed differences could be interpreted as intermittent reconnection in different loop threads that have different plasma and magnetic environments.
The remote brightening away from the reconnection region is found in the tracks of the fast propagating ``flows''.
This brightening seems to be a secondary activity triggered by the reconnection event, and it
also produces a slow flow with a speed of 64\,\kms, comparable to the slow flows produced in the reconnection site.
If we understand the fast propagating ``flows'' as shock flows, the remote brightening could be a response of the exhaust region\,\citep{2005JGRA..110.1107G,He2017inprep} and the flow could be the result of heating therein.
This, however, cannot be confirmed with the present data.

\par
Beside the apparent flows, the reconnection has also flared up the loops that could be seen in all AIA channels.
In Figure\,\ref{fig:prflow}c, we present the light-curves taken from a small region crossing the flaring loops.
The peaks in the light-curves between 07:12\,UT and 07:30\,UT are representative of the activities resulted from the reconnection event.
Multiple peaks are clearly present in the light-curves of the passbands of 1600\,\AA, 304\,\AA, 171\,\AA\ and 193\,\AA.
Between any two passbands, these peaks in their light-curves do not show any detectable delay.
Why are these light-curves well in-phase?
One interpretation could be multi-thermal nature of the phenomenon.
Alternatively,  since the bandwidths of the AIA passbands are relatively large, these peaks could be mainly contributed by the continuum emission because of the density increase resulting from the running blobs.
These peaks are also corresponding to small humps in the AIA 131\,\AA\ and 94\,\AA\ light-curves, which are also contributed by continuum emissions.
In this period of time, the AIA 131\,\AA\ and 94\,\AA\ light-curves are closed to be single-peaked, especially the later one.
This suggests that the intensity increase from the running blobs might not be the only major contributor 
and the thermal/heating effect could not be neglected.
The major peak in the 94\,\AA\ light-curve presents at 07:22:24\,UT, 
while a small hump present in the 131\,\AA\ light-curve presents at 07:21:19\,UT.
Since the 131\,\AA\ passband has large contribution from Fe\,{\sc viii} (0.4 MK) emission\,\citep{2010A&A...521A..21O}.
it suggests that this passband is representative of cooler temperature than that of the 94\,\AA\ (Fe\,{\sc xviii}, 6.3 MK).
This time delay (65\,s) between the emissions in two passbands could be a result of the heating process.

\par
The light-curves and the online animation also reveal that the AIA\,94\,\AA\ brightness of the coronal loops remain in a higher level for a relatively-long period of time.
It suggests that the post-reconnection loops are relatively stable in temperature around 6.3 MK.
This is consistent with the visibility of the coronal loops in the AIA\,94\,\AA\ channel during the pre-reconnection stage.
Most plausibly, the brightening in the post-reconnection loops results from the heated filamentary plasmas that have been filled in the loops.
Therefore, such reconnection processes could be a considerable approach feeding mass and energy into coronal loops.

\section{Conclusion}
\label{sect_concl}
In this study, we present high-resolution observations of a magnetic reconnection event taken with NVST and SDO/AIA/HMI.
The magnetic reconnection occurred between coronal loops and threads of a twisted arch filament system (AFS).
The NVST \halpha\ images reveal that the inflows of the reconnections were driven by relaxation of the AFS.
The untwined motions of the twisted AFS result in some of its threads encountering the coronal loops and
providing inflows of the reconnection.
This could be an approach how random motions in the photosphere drive magnetic reconnection in higher solar atmosphere.
The untwined motions of the AFS threads have an apparent acceleration of $\sim$15\,m\,s$^{-2}$, and an apparent speed of 5.5\,\kms\ is found for the motions of the AFS threads encountering the coronal loops.
Our observations reveal that the twisted magnetic system could release its free energy into the upper solar atmosphere through reconnection processes.

\par
The observations reveal the rearrangement of connectivities in the region due to the reconnection event.
An elongated current-sheet-like feature with 3.5\arcsec\ in length and 0.5\arcsec\ in width is also observed with the \halpha$+$1\,\AA\ images.
Small-scale magnetic cancelation is present below the inferred reconnection site.
The reconnection heats the filamentary plasma and allows them to be released into the coronal loops.
The flows along the loops driven by the reconnection appear to have various speeds in a range of about 40--80\,\kms, possibly result from intermittent reconnection.
The emission increase in the coronal loops could be results of the filling-in filamentary plasma and/or heating.
Therefore, these observations also suggest that reconfiguration of field lines in a solar magnetic system could result in transferring of mass between different loops and induce heating therein.

\acknowledgments
{\it Acknowledgments:}
We thank the anonymous referee for his/her comments.
This research is supported by
National Natural Science Foundation of China (41404135, 41474150 and 41474149).
ZH thanks the China Postdoctoral Science Foundation and the Young Scholar Program of Shandong University, Weihai (2017WHWLJH07).
We acknowledge many discussions during the magnetic reconnection workshop in the Purple Mountain Observatory.
We thank the FSO staff for their help and hospitalities during our stays in the observatory.
Courtesy of NASA/SDO, the AIA and HMI science teams and JSOC.

\bibliographystyle{aasjournal}

\end{document}